\documentclass[referee]{jfm}

\usepackage{graphicx}
\usepackage{natbib}

\usepackage{graphicx,rotating,soul}
\usepackage[normalem]{ulem}
\usepackage{amssymb,amsmath,amsbsy}
\usepackage{epsf}
\newcommand{\mb}[1]{\mbox{\boldmath$#1$}}
\usepackage{subfigure}
\usepackage{appendix}
\usepackage{natbib}
\usepackage{color}
\title[Metachronal beating of magnetic artificial cilia]{Microfluidic propulsion by the metachronal beating of magnetic artificial cilia: a numerical analysis} 

\author[S. N. Khaderi, J. M. J. den Toonder and P. R. Onck]{S.\ns N.\ns K\ls H\ls A\ls D\ls E\ls R\ls I$^1$,\ns J.\ns M.\ns J.\ns D\ls E\ls N\ns T\ls O\ls O\ls N\ls D\ls E\ls R$^2$\ns \and P.\ns R.\ns O\ls N\ls C\ls K\ls$^1$\footnote{Corresponding author: p.r.onck@rug.nl}\ns}
\affiliation{$^1$Zernike Institute for Advanced Materials, University of Groningen, Groningen, The Netherlands.\\[\affilskip]
$^2$Eindhoven University of Technology, Eindhoven, The Netherlands.
}%

\begin{document}

 \maketitle
\begin{abstract}
In this work we study the effect of metachronal waves on the flow created by magnetically-driven plate-like artificial cilia in microchannels using numerical simulations.  The simulations are performed using a coupled magneto-mechanical solid-fluid computational model that captures the physical interactions between the fluid flow, ciliary deformation and applied magnetic field.  When a rotating magnetic field is applied to  super-paramagnetic artificial cilia, they mimic the asymmetric motion of natural cilia, consisting of an effective and recovery stroke. When a phase-difference is prescribed between neighbouring cilia, metachronal waves develop. Due to the discrete nature of the cilia, the metachronal waves change direction when the phase difference becomes sufficiently large, resulting in antiplectic as well as symplectic metachrony. We show that  the fluid flow created by the artificial cilia is significantly enhanced in the presence of metachronal waves and that the fluid flow becomes unidirectional. Antiplectic metachrony is observed to lead to a considerable enhancement in  flow compared to symplectic metachrony, when the cilia spacing is small. Obstruction of flow in the direction of the effective stroke for the case of symplectic metachrony was found to be the key mechanism that governs this effect.

\end{abstract}

 

\section{Introduction}

The control of fluid flow in  channels of micron-scale dimensions is essential for proper functioning of any lab-on-a-chip device. The fluid transport in microchannels is often performed by downscaling conventional methods such as syringe pumps, micropumps \cite[]{0960-1317-14-6-R01, JeonN.L._la000600b, SchillingE.A._ac015640e}, or by exploiting electro-magnetic fluid manipulation principles, as in electro-osmotic \cite[]{lingxin,shulin} and magneto-hydrodynamic \cite[]{jonathan_west} devices. In search for novel ways to propel fluids at micron scales, we let nature be our guide. Nature uses hair-like structures, called cilia, attached to the surfaces of micro-organisms, to propel fluids at small length scales. The typical length of a cilium is 10 microns. Cilia beat in a whip-like asymmetric manner consisting of an effective stroke and a recovery stroke. Moreover, when many cilia operate together, hydrodynamic interactions cause them to beat out-of-phase \cite[]{ShayGueron06101997}, leading to  the formation of metachronal waves, and an enhanced fluid flow  \cite[]{satir_sale_annual_review}. The specific metachrony is termed symplectic (or antiplectic) when the  metachronal wave is in the same (or opposite) direction as the effective stroke. The cilia on a Paramecium exhibit antiplectic metachrony, whereas the cilia on Opalina exhibit symplectic metachrony   \cite[]{blake_sleigh_review}.  The asymmetric motion of natural cilia is due to the intricate interaction between the cilia micro-structure (axoneme) and the internal driving force generated by ATP-enabled conformational changes of the motor protein dynein. It is a challenging task to design the artificial counterpart of natural cilia, by using external force fields for actuation in order  to mimic the asymmetric motion of natural cilia. An early attempt to create artificial cilia was based on electrostatic actuation  of arrays of plate-like artificial cilia \cite[]{den_toonder}. Although effective flow and mixing were achieved, movement of these artificial cilia was not asymmetric as in the case of natural cilia. It was predicted using numerical simulations that an array of identical super-paramagnetic or permanently magnetic two-dimensional plate-like cilia can mimic the planar asymmetric motion of natural cilia when exposed to a uniform magnetic field  \cite[]{khaderi}. These magnetic plate-like cilia can be realised, for instance,  by using polymer films with embedded  super-paramagnetic (or permanently magnetic)  nano-particles \cite[see e.g.~][]{3dexpt_cilia, belardi_tolouse, schorr_tolouse}. In contrast with the plate-like cilia, rod-like structures that mimic the three-dimensional motion of nodal cilia to create fluid propulsion have also been fabricated \cite[]{self_assembles_cilia, Shields, evans_cilia}. In \cite{katz_walker},  a novel method of fluid propulsion based on magnetic walkers was presented. Artificial cilia based on photo-actuation have also been realised in the recent past \cite[]{printed_cilia}.

In  previous numerical studies we focused on the flow created by an array of synchronously-beating plate-like  cilia whose motion is planar and  asymmetric, in the absence \cite[]{khaderi} and presence of fluid inertia \cite[]{khaderi_inertia}.  It was reported that a substantial but fluctuating flow is created in the former, while in the latter  the flow increases significantly  as the Reynolds number is increased. In addition, the fluid flow can become unidirectional in the presence of fluid inertia. In this work we explore another aspect of natural ciliary propulsion using numerical simulations -  the metachronal motion of cilia, by allowing the asymmetrically-beating artificial cilia to move  out-of-phase.  The out-of-phase motion of the cilia is achieved by applying a magnetic field that has a phase lag between adjacent cilia. The existing literature on the metachronal motion of natural cilia could provide insights on the flow generated in the presence of metachronal waves.

In the case of natural cilia the metachronal motion  is analysed principally  for two reasons. First, to find the effect of the metachronal waves on the flow created and second, to find the physical origin of the metachronal waves. Theoretical and numerical studies have been undertaken by biologists and fluid mechanicians to understand the  flow created by an array of cilia  \cite[see for e.g.~the reviews by][]{brennen_cilia_flagella, blake_sleigh_review,  smith_modelling_review}. Most of these analyses have been performed to model the flow of specific biological systems (e.g.~micro-organisms or airway cilia), however, a systematic study is lacking.    In the following, we outline a number of studies in which the effect of the metachronal waves on fluid transport has been studied. Modelling approaches to understand the cilia-driven flow include  the envelope model \cite[]{brennen_cilia_flagella, blake_envelope_1971, blake_spherical_envelope}, the sublayer model \cite[]{blake_sublayer_1972, ShayGueron06101997, smith_discrete_cilia, liron_cilia_between_plates, gauger_cilia, ShayGueron_energy}, fluid structure interaction models using a lattice-Boltzmann approach \cite[]{kim_netz}, and the immersed boundary method \cite[]{Dauptain}. 
In the envelope model, the cilia are assumed  to be very densely spaced so that the fluid experiences  an oscillating surface consisting of the tips of the cilia. The envelope model is accurate only when the cilia are spaced very close together, which has  only been observed in the case of symplectic metachrony \cite[]{blake_envelope_1971, blake_spherical_envelope}. 
In the sublayer model \cite[]{blake_sublayer_1972}, the cilia are represented by a distribution of Stokeslets with appropriate mirror images to satisfy the no-slip condition on the surface to which the cilia are attached. The sublayer model  predicts that for an organism that exhibits antiplectic metachrony, the flow created  is lower than for cilia beating in-phase. In the case of an organism exhibiting symplectic metachrony, the opposite trend is observed.
In  the numerical study of \cite{gauger_cilia}, the flow due to the out-of-phase motion of a finite number of magnetic cilia subjected to an oscillating external magnetic field was studied. The magnetic cilia generate an asymmetric motion due to the difference in the speed of oscillation of the magnetic field during  the effective and recovery strokes. In contrast to  \cite{blake_sublayer_1972},  it was predicted that the flow in the case of antiplectic metachrony is larger than the flow created by a symplectic metachrony for a particular inter-cilia spacing. 

Early experiments indicated that the hydrodynamic coupling between cilia could be the cause for the formation of the metachronal waves \cite[see for e.g.~the review by][]{kinosita_review}. By mimicking the ciliary motion of Paramecia using an internal actuation mechanism, it was demonstrated that cilia, which were initially beating in-phase, will form an antiplectic metachronal wave after a few beat cycles \cite[]{ShayGueron06101997}. This behaviour was explained to be an outcome of the hydrodynamic interactions between neighbouring cilia. Similar hydrodynamically-caused metachronal motion of the cilia was also observed in the numerical work of \cite{mitran}.  In \cite{ShayGueron_energy},  it was reported that  in the presence of the metachronal wave the cilia become more efficient in creating flow.  The synchronization and phase locking of the cilia have also been analysed  using simple experimental \cite[]{qian_syn} and analytical \cite[]{niedermayer, vilfan_synchronization} models. It was found that some degree of flexibility is required for the phase locking of the cilia to take place \cite[]{niedermayer,qian_syn}. The requirement of the flexibility for synchronization is also confirmed from the more detailed model of \cite{kim_netz}. In the aforementioned studies, however, the metachronal wave is an outcome of that specific system, and the flow or the efficiency has not been studied for different types of metachronal waves.

The goal of this paper is, therefore, to obtain a full understanding of the dependence of flow on the magnetically-induced out-of-phase motion of an array of asymmetrically beating plate-like artificial cilia at low Reynolds numbers. We will  answer the following questions using a coupled solid-fluid magneto-mechanical computational model.  How does the generated flow in the presence of metachrony differ from the flow generated by cilia that beat in-phase? How does the flow depend on the metachronal wave speed and its direction, and how  does it depend on the cilia spacing? We answer these questions in the light of  magnetic artificial cilia which exhibit an asymmetric motion and beat out-of-phase. However, the results are equally applicable to any ciliary system in which  the cilia exhibit an asymmetric and out-of-phase motion. 

The paper is organised as follows. The boundary value problem, the governing equations and the numerical solution methodology are explained in section \ref{sec:problem_cilia_meta_delta_phi}.  In section \ref{sec:results_cilia_meta_no_bf}, the  physical mechanisms responsible for the enhanced flow in the presence of metachronal waves are discussed. The quantitative variation of the flow as a function of the phase difference and cilia spacing is given. Finally, the outcome of the analysis is summarised in section \ref{sec:conclusion_out-of-phase}.

\section{Problem statement and approach}\label{sec:problem_cilia_meta_delta_phi}
\begin{figure}
	              \centering
		      \subfigure[]{\includegraphics[width=8cm]{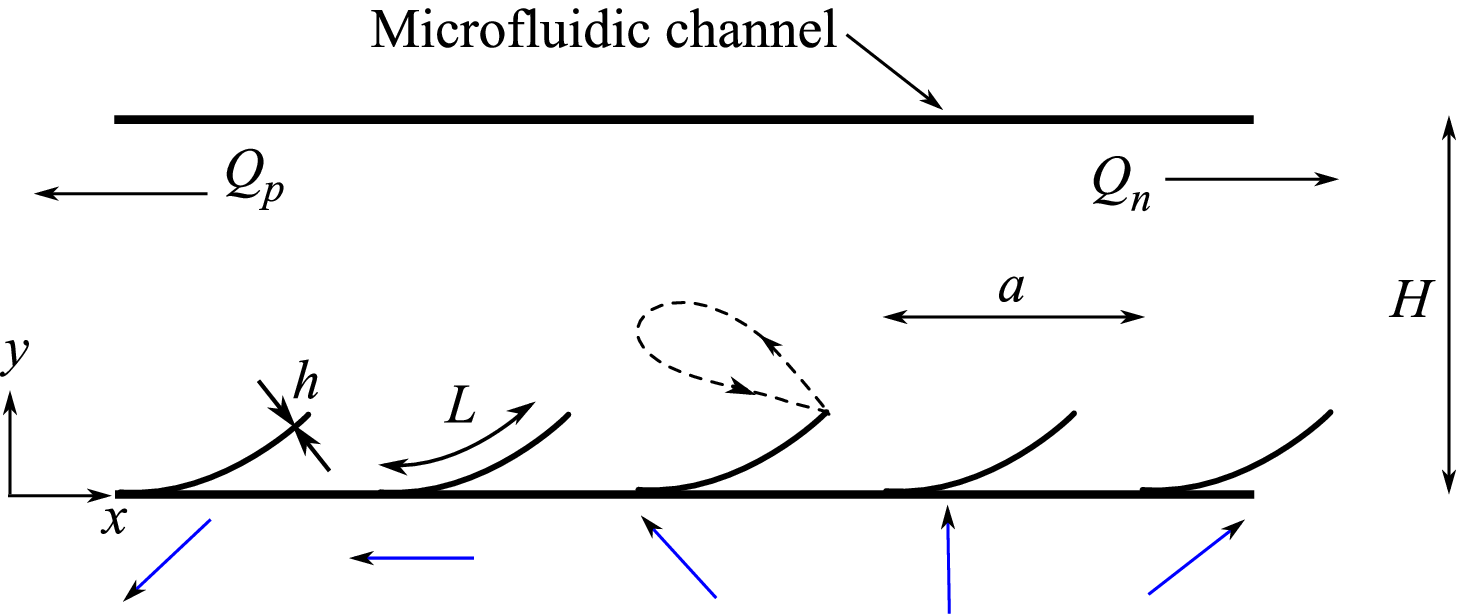}}\ \ \ \ \ \ \ 
		      \subfigure[]{\includegraphics[width=3cm]{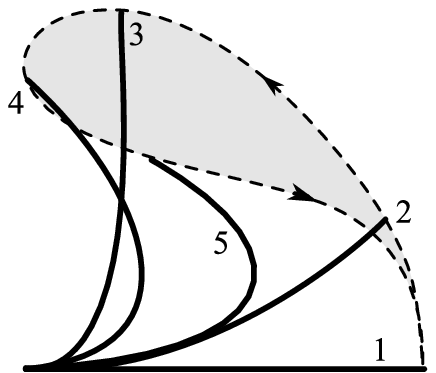}}
		      \caption{(Color online) (a) Schematic representation of the problem analysed. We study an infinitely long microfluidic channel consisting of equal-sized cilia (having length $L$ and thickness $h$) spaced a distance $a$ apart. The variation of magnetic field in space is shown using blue arrows.   $Q_p$ and $Q_n$ denote the flow in the direction of  the effective and recovery stroke, respectively. (b) Typical asymmetric motion of a cilium. The  dashed lines  represents   the trajectory of the tip of an individual cilium.}
                      \label{fig:case3_meta_schematic}
	  \end{figure}

We study the flow in an infinitely long channel of height $H$ created by a   two-dimensional array of plate-like magnetic artificial cilia (having length $L$ and thickness $h$), which are actuated using a rotating magnetic field which is uniform over each cilium, but with a phase difference between  adjacent cilia. The external  magnetic field experienced by the  $i^\text{th}$ cilium is
\begin{equation}\label{eq:non_uniform_field}
 B_{xi}=B_0 \cos(\omega t - \phi_i), \ \ \ \  B_{yi}=B_0 \sin(\omega t - \phi_i),   
\end{equation}
where $B_0$ is the magnitude of the applied magnetic field, the phase of the magnetic field $\phi_i=2\pi (i-1)/n$, $\omega=2\pi / t_\text{ref}$ is the angular frequency and $t_\text{ref}$ is the time period of rotation of the magnetic field. The magnetic field experienced by the individual cilia during a particular instance in time is shown using the blue arrows in Fig.~\ref{fig:case3_meta_schematic}(a). The phase difference in the applied magnetic field between adjacent cilia is $\Delta\phi=2\pi/n$. The chosen form of the phase $\phi_i$ makes the phase of the magnetic field at every $n^\text{th}$ cilium identical. That is, the magnetic field is periodic after $n$ repeats of cilia.
Consequently,  the applied magnetic field travels $n$ cilia units in time $t_\text{ref}$, so that the phase velocity of the magnetic field is $n/t_\text{ref}=\omega/\Delta\phi$ (in cilia per second). The phase velocity is to the right (positive) and the magnetic field at each cilium position rotates counterclockwise with time. The typical asymmetric motion of a cilium is shown in Fig.~\ref{fig:case3_meta_schematic}(b).
The cilia are tethered at one end to the  surface, while the other end is free. The trajectory of the free end of a  typical cilium is represented by the dashed lines in Fig.~\ref{fig:case3_meta_schematic}(b), with the arrows representing the direction of motion. 

Due to the super-paramagnetic (SPM) nature of the cilia, for which the magnetization is proportional to the magnetic field,  the magnetic body couple ($\mb{N}=\mb{M}\times\mb{B}_0$, where $\mb{M}$ is the magnetization of the cilia and $\mb{B}_0=(B_{x}, B_{y})$ is the magnetic field experienced by the cilia)  depends only on the orientation and magnitude of the magnetic field, but not on its sign. As a result, the body couple at the $i^\text{th}$ cilium $N_{zi}$, which determines its motion,  scales with $ \sin\left( 2\omega t -    2 \phi_i \right) $ \cite[]{CambridgeJournals:431278}.  This has consequences for the motion of the cilia, both  temporally and spatially. Temporally, the frequency of the magnetic couple is twice that of the applied magnetic field. This results in  two cilia beats for one 360$^\circ$ rotation of the magnetic field. Spatially, the  phase  of the magnetic couple is  twice that of the applied magnetic field, so that the phase difference between neighbouring cilia is twice as large.  This means that the magnetic couple is periodic after $n/2$ cilia. Since both the frequency and phase difference increase by a factor 2, the phase velocity of the magnetic torque remains equal to that of the magnetic field, i.e.~$\omega/\Delta\phi$. Note, however, that the phase velocity of the magnetic torque is equal to the velocity of the metachronal wave (i.e., the actually observed deformational wave travelling over the cilia) only when the phase difference $\Delta\phi$ is small (i.e.~$n$ is large).

\begin{figure}
	              \centering
 \includegraphics[width=7 cm]{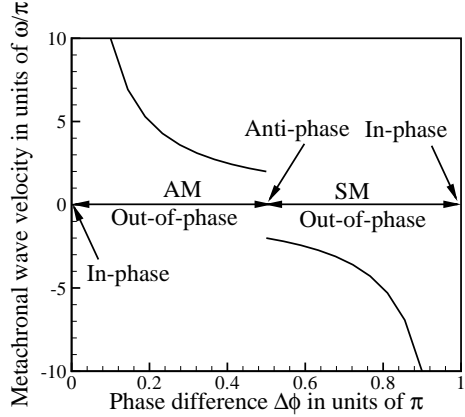}
\caption{Metachronal wave velocity as a function of the phase difference $\Delta\phi$ in the magnetic field between adjacent cilia. AM and SM refer to antiplectic and symplectic metachrony respectively.}
\label{fig:case3_meta_apparent_wave_velocity}
\end{figure}

When the phase difference is too large, the metachronal wave can change sign, so that the metachronal wave is observed to travel in a direction opposite to the direction of the magnetic field (see  appendix \ref{sec:apparent_velocity}).   The metachronal wave  velocity is equal to $\omega/\Delta\phi$ (i.e.~to the right) when $0<\Delta\phi<\pi/2$, and it is equal to  $-\omega/(\pi-\Delta\phi)$ (i.e.~to the left) when $\pi/2<\Delta\phi<\pi$, see Fig.~\ref{fig:case3_meta_apparent_wave_velocity}. When $\Delta\phi= 0$, the magnetic couple is uniform and all cilia beat in-phase. When $\Delta \phi =\pi$, the magnetic couple acting on two neighboring cilia is the same (because the phase difference of the magnetic couple is $2\Delta\phi=2\pi$), and again, all the cilia beat in-phase.  When $\Delta \phi=\pi/2$, the positive metachronal wave velocity is equal in magnitude to its negative counterpart. In such a condition, a standing wave is observed which causes  the adjacent cilia to move in anti-phase. 
When $0<\Delta \phi<\pi/2$ the metachronal wave velocity is positive, i.e.~to the right in Fig.~\ref{fig:case3_meta_schematic}. Consequently, the metachronal wave velocity is opposite to the direction of the effective stroke, which is commonly addressed as antiplectic metachrony (AM). When  $\pi/2<\Delta \phi<\pi$, the metachronal wave velocity is in the same direction as the effective stroke and is referred to as symplectic metachrony (SM), see Fig.~\ref{fig:case3_meta_apparent_wave_velocity}.

\subsection{Governing equations}
We now briefly discuss the coupled solid-fluid magneto-mechanical numerical model used to study fluid propulsion using magnetically actuated plate-like artificial cilia.
In typical microfluidic channels the height $H$ is smaller than the out-of-plane width. Moreover, the artificial cilia under study are plate-like (having an out-of-plane width $b$ much larger than their thickness $h$ and length $L$) exhibit a planar beat motion. Therefore, any variation in the out-of-plane direction can be neglected and under these assumptions it is sufficient to model the artificial cilia and the resulting flow in a two-dimensional setting. 
\subsubsection{Solid dynamic model}\label{sec:solids_eom}
We model the cilia as elastic Euler-Bernoulli beams taking into consideration geometric non-linearity in an updated Lagrangian framework.  
As a starting point for the Euler-Bernoulli beam element formulation we use the principle of virtual work \cite[]{malvern} and equate the virtual work of the external forces at time $t+\Delta t$ ($\delta W_\text{ext}^{t+\Delta t}$) to the internal work ($\delta W_\text{int}^{t+\Delta t}$).
The internal virtual work is given by
\begin{equation}
 \delta W_\text{int}^{t+\Delta t}= \int_V\left( \sigma \delta\epsilon + \rho(\ddot{u}\delta u+\ddot{v}\delta v)\right)dV,
\end{equation}
where $u$ and $v$ are the axial and transverse displacements of a point on the beam and $\rho$ is the density of the beam. Furthermore,  $\sigma$ is the axial stress and $\epsilon$ is the corresponding strain, given by
\begin{eqnarray}
 \epsilon&=&\frac{\partial u}{\partial x}+\frac{1}{2}\left(\frac{\partial v}{\partial x} \right)^2 -y\frac{\partial^2 v}{\partial x^2}\nonumber.
\end{eqnarray}
The external virtual work is
\begin{equation}\begin{split}\label{eqn:solids_external_virtual_work}
     \delta W_\text{ext}^{t+\Delta t}=&\int \left(f_x \delta u+f_y \delta v+ N_z \frac{\partial \delta v}{\partial x}  \right)Adx \\
     &+ \int \left(t_x\delta u+t_y\delta v\right)b dx,
\end{split}\end{equation}
where $f_x$ and $f_y$  are the magnetic body forces in the axial and transverse directions, $N_z$ is the magnetic body couple in the out-of-plane direction, $t_x$ and $t_y$ are the surface tractions  and $b$ is the out-of-plane thickness of the cilia.

We follow the approach used in \cite{ratna_jmps} to linearise and discretise the principal of virtual work to get,
\begin{equation}\label{eq:after_linearization}\begin{split}
 \delta\mb{p}^T\left(\mb{K}\Delta\mb{p}+\mb{M}\ddot{\mb{p}}^{\text{t}+\Delta \text{t}}-\mb{F}^\text{t+$\Delta $t}_\text{ext}+\mb{F}^\text{t}_\text{int}\right)=0,
 \end{split}\end{equation}
 where $\mb{K}$ is the stiffness matrix that combines both material and geometric contributions, $\mb{M}$ is the mass matrix that can be found in \cite{cook_fem}, $\mb{F}^\text{t+$\Delta $t}_\text{ext}$ is the external force vector,  $\mb{F}^\text{t}_\text{int}$ is the internal force vector, $\Delta \mb{p}$ is the nodal displacement increment vector and $\ddot{\mb{p}}$ is the nodal acceleration vector. The nodal acceleration vector is discretized in time using Newmark's algorithm (using parameters $\gamma=1.0$ and $\beta=0.5$) so that Eqn.~\ref{eq:after_linearization} can be written in terms of the velocity of the beam.  The complete discretized equations of motion for the solid mechanics model can be found elsewhere \citep{khaderi}.

 \subsubsection{Magnetostatics}\label{sec:magnetostatics}
To find the resulting magnetic forces, the magnetization of the cilia has to be calculated by solving the Maxwell's equations in the deformed configuration at every time increment.  The Maxwell's equations for the magnetostatic problem with no external currents are
\begin{equation}\label{eqn:maxwell_equations}
     \mb{\nabla}\cdot\mb{B}=0 \ \ \ \   \mb{\nabla}\times\mb{H}=0,
\end{equation}
with    the constitutive relation $\mb{B}=\mu_0(\mb{M}+\mb{H}),$
where $\mb{B}$ is the magnetic flux density (or magnetic induction), $\mb{H}$ is the magnetic field, $\mb{M}$ is the magnetization, and $\mu_0$ is the permeability of vacuum. Equation \ref{eqn:maxwell_equations} is solved for $\mb{M}$ and $\mb{B}$ using the boundary element method \cite[]{khaderi}. The magnetic couple per unit volume is given by $\mb{N}=\mb{M}\times \mb{B}_0$. As the simulations are two dimensional,  the only non-zero component of magnetic body couple is $N_z$ which is the source for the external virtual work in Eqn.~\ref{eqn:solids_external_virtual_work}. Since the applied magnetic field is uniform for each cilium, the magnetic body forces due to field gradients are absent.

\subsubsection{Fluid dynamics and solid fluid coupling}
We study the flow created by artificial cilia in the limit of low Reynolds number. The fluid is assumed to be Newtonian and incompressible. The physical behaviour of the fluid is governed by the Stokes equation:
\begin{equation}\label{eq:stokes}
  \begin{split}
 -\mb{\nabla }p+2\mu\mb{\nabla} \cdot \mb{D}&=0,\\
   \mb{\nabla}\cdot\mb{u}&=0,
 \end{split}
\end{equation}
where $p$ is the   pressure in the fluid, $\mb{D}$ is the rate of deformation tensor, $\mb{u}$ is the velocity of the fluid and $\mu$ is the viscosity of the fluid. 
The set of equations in Eqn.~\ref{eq:stokes} is solved using Eulerian finite elements based on the Galerkin method. The fluid domain is discretized into quadrilaterals in which the velocity and pressure of the fluid are interpolated quadratically and linearly, respectively. The velocity is calculated at the vertices, mid-sides and mid-point of the quadrilateral, and the pressure is calculated at the vertices.   The solid and fluid domains are coupled by imposing the constraint that the velocity  at the nodes of  the solid beam are equal to the velocity of the  surrounding fluid (point collocation method).  This coupling is established with the help of Lagrange multipliers  using the fictitious domain method. Details of the Eulerian finite element model and the coupling procedure can be found in \cite{vanloon}. 

The fluid domain used for the simulations has a width $W$ and height $H$ (Fig.~\ref{fig:mesh_meta}). For each value of $a/L$, we choose $n$ to be a fraction $p/q$ larger than 2, with $p$ and $q$ integers, yielding a range of phase differences $\Delta\phi=2\pi/n$ between 0 and $\pi$.  For each value of $p/q$, a unit-cell of width $W=pa$ needs to be chosen to account for periodicity in the magnetic couple, unless $p$ is an even integer, for which $W=pa/2$ suffices.   For example, let $p=10$ and $q=3$. Now, $n=10/3$ and the phase difference $\Delta\phi$ is equal to $3\pi/5$. To maintain periodicity in the magnetic couple, the width of the unit-cell should be $5a$ (containing 5 cilia).  The top and bottom of the unit-cell are the channel walls, on which no-slip boundary conditions are applied,
\[ \mb{u}_\text{top}=\mb{u}_\text{bottom}=\mb{0},  \]
while the left and right ends are periodic in velocity
\[ \mb{u}_\text{left}=\mb{u}_\text{right}.  \]

\subsubsection{Solution procedure} 
\begin{figure}\centering
\includegraphics[width=12 cm]{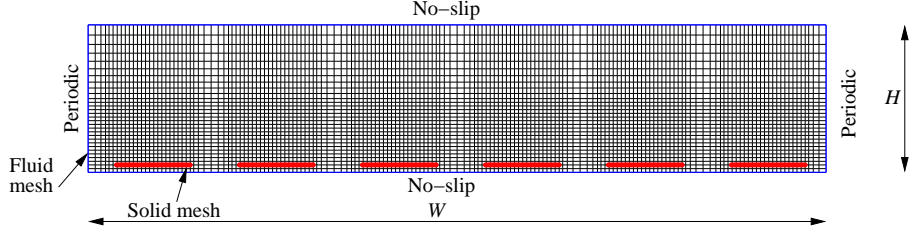}
\caption{Fluid (black) and solid (red) mesh used for the simulations. The mesh corresponds to $\Delta\phi=\pi/6$ and $a=1.67L$.}
\label{fig:mesh_meta}
\end{figure}

The solution procedure is as follows. The Maxwell's equations are solved at every time instant to solve for the magnetic field. From the magnetic field, the magnetic body couple acting on the cilia is calculated and is provided as an external load to the coupled solid-fluid model, which simultaneously solves for the cilia velocity, and the velocity and pressure of the fluid. The velocity of the cilia is integrated using Newmark's algorithm to obtain its new position, and the procedure is repeated. Each cilium is discretized into 40 elements and every fluid domain of size $a\times H$ is discretized into $28 \times 30$ elements, with the mesh being refined near each cilium. A typical mesh used for the simulations is shown in Fig.~\ref{fig:mesh_meta}. A fixed time-step of 1 $\mu$s was used for all the simulations reported in this paper.  The spatial and temporal convergence of the numerical model is discussed in appendix \ref{sec:convergence}.  The particles and streamlines are obtained from the velocity field in the fluid using the visualization software Tecplot \cite[]{tecplot}. Also here care should be taken to accurately resolve the velocity field.

\subsection{Parameter space}
The physical dimensionless numbers that govern the behavior of the system are the magneto-elastic number   $M_n=12 B_0^2L^2/\mu_0Eh^2$ - the ratio of the magnetic to the elastic forces, the fluid number $F_n=12\mu L^3/Eh^3t_\text{beat}$ - the ratio of viscous forces acting on the cilia to the elastic forces, and the inertia number $I_n=12\rho L^4/Eh^2t_\text{beat}^2$ - the ratio of the inertia forces of the cilium to its elastic forces, \cite[see][]{khaderi}. Here, $E$ is the elastic modulus of the cilia, $h$ is the thickness, $\rho$ is the density of the cilia, $\mu$ is the fluid viscosity, $t_\text{beat}(=t_\text{ref}/2)$ is the time period of one beat cycle and $\mu_0$ is the magnetic permeability. The geometric parameters that govern the behavior of the system are  the phase difference $\Delta \phi$, the cilia spacing $a$, their length $L$ and the height of the channel $H$. We  study the flow created as a function of the cilia spacing $a$ (normalised with the length $L$)  and the phase difference $\Delta\phi$ for the following set of parameters: $F_n =0.15$, $M_n=12.2$, $I_n=4.8\times 10^{-3}$ and $H/L=2$. The values of the physical parameters correspond to $L =$ 100 microns, $E = 1$ MPa, the thickness of cilia being $h=2\ \mu$m at the fixed end and 1 $\mu$m at the free end, $\rho= 1600$ kg/m$^3$, $\mu= 1$ mPas, $B_0=22.6$ mT and the cycle time $t_\text{ref} =20$ms. The magnetic susceptibilities of the cilia are 4.6 along the length and 0.8 along the thickness \cite[]{Rijsewijk}. 

The fluid propelled is characterised by two parameters: the net volume of the fluid transported during a ciliary beat cycle  and the effectiveness. The horizontal velocity field in the fluid at any $x$ position, integrated along the channel height gives the instantaneous flux through the channel. This flux integrated in time over the effective and recovery stroke gives the positive ($Q_p$) and negative ($Q_n$) flow, respectively (see Fig.~\ref{fig:case3_meta_schematic}). Due to the asymmetric motion, the positive flow is larger  than the negative flow, generating a net area flow per cycle ($Q_p-Q_n$) in the direction of the effective stroke.  
The effectiveness,  defined as $(Q_p-Q_n)/(Q_p+Q_n)$, indicates which part of the totally displaced fluid is effectively converted into a net flow. An effectiveness of unity represents a unidirectional flow.

\section{Results and discussion }\label{sec:results_cilia_meta_no_bf}
\begin{figure}
\begin{minipage}{0.6\textwidth }
  \centering
               \subfigure[\ $t=0$]{\includegraphics[scale=1.0]{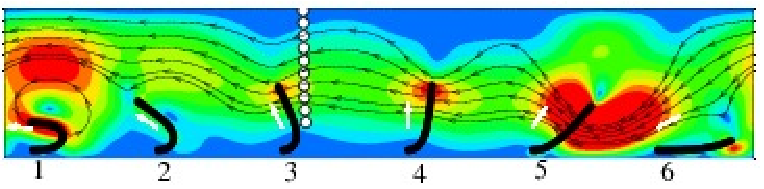}\label{fig:n_6_wk_1_particle_0p03088}}\\
\subfigure [\ $t=t_\text{beat}$/6]{\includegraphics[scale=1.0]{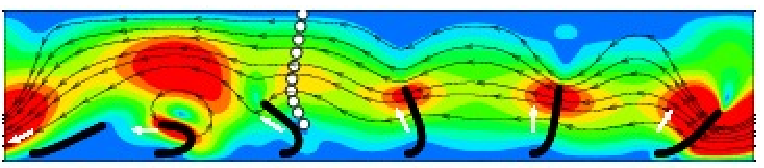}\label{fig:n_6_wk_1_particle_0p032}}\\
\subfigure[\ $t=2t_\text{beat}/6$]{\includegraphics[scale=1.0]{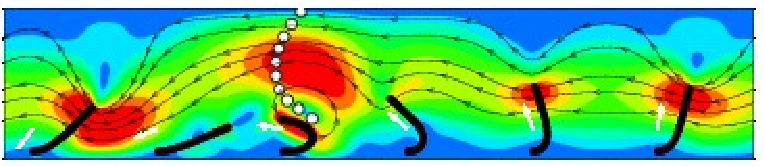}\label{fig:n_6_wk_1_particle_0p0335}}\\
\subfigure[\ $t=3t_\text{beat}/6$]{\includegraphics[scale=1.0]{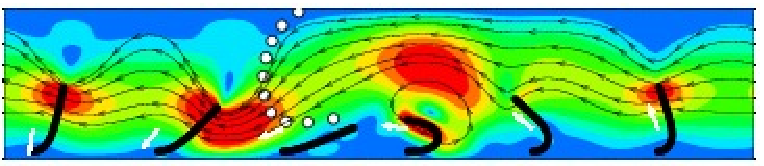}\label{fig:n_6_wk_1_particle_0p03538}}\\
\subfigure[\ $t=4t_\text{beat}/6$]{\includegraphics[scale=1.0]{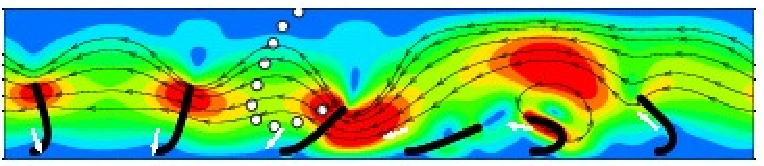}\label{fig:n_6_wk_1_particle_0p03688}}
\subfigure[\ $t=5t_\text{beat}/6$]{\includegraphics[scale=1.0]{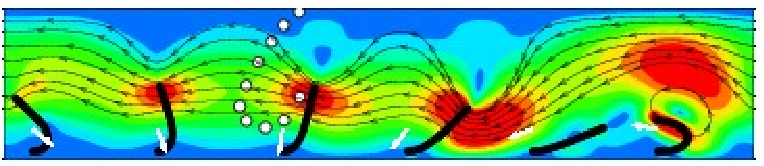}\label{fig:n_6_wk_1_particle_0p03888}}

\end{minipage}
\begin{minipage}{0.1\textwidth}
  \subfigure[]{\includegraphics{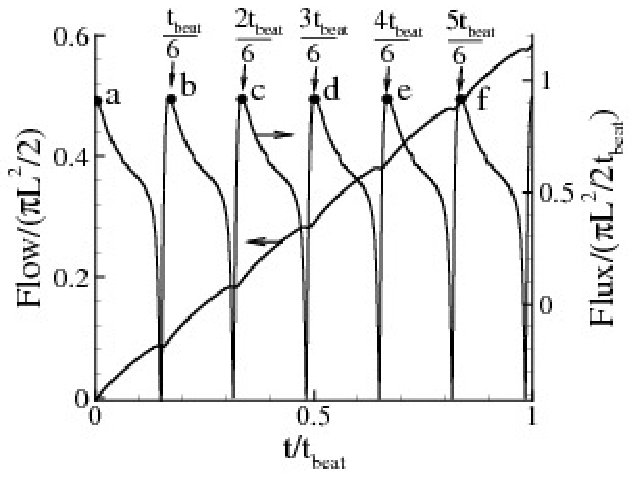}\label{fig:case3_meta_flux_time_n6}}
\includegraphics{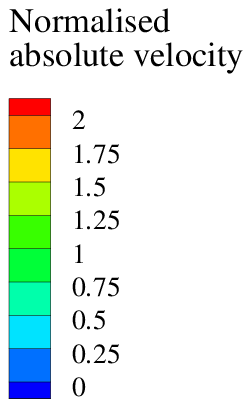}
\end{minipage}
    \caption{(Color online) (a)-(f) Out-of-phase motion of cilia during a representative cycle for $\Delta\phi=\pi/6$ ($n=12$) with the wave moving to the  right (antiplectic metachrony) for $a/L=1.67$.   The contours represent the absolute velocity normalised with $L/t_\text{beat}$.  The direction of the velocity is represented by streamlines. The white circles represent fluid particles. The applied magnetic field at each cilium is represented by the white arrows. (g) Instantaneous  flux  (right axis) and flow (or accumulated flux, left axis) as a function of time with the instants  (a)-(g) duly marked.     }
     \label{fig:case3_meta_particle_tracking}
\end{figure}

To obtain  an understanding of fluid flow due to the out-of-phase motion of cilia, we analyse the  case of antiplectic metachrony with a phase difference $\Delta\phi=2\pi/n=2\pi/12$.  Since $n$ is  even, a unit-cell of width $6a$ consisting of 6 cilia is chosen, see Fig.~\ref{fig:case3_meta_particle_tracking}.
The contours represent the absolute velocity normalised with $L/t_\text{beat}$. The direction of the velocity field can be determined from the arrows on the streamlines. The white arrows represent the applied magnetic field for each cilium. Animations of the ciliary motion for the cases of symplectic, antiplectic and anti-phase motion are provided as  supplementary information.

The snapshots shown in Figs.~\ref{fig:n_6_wk_1_particle_0p03088}-\ref{fig:n_6_wk_1_particle_0p03888} correspond to the time instances when the flux generated by the cilia is maximum. In Fig.~\ref{fig:case3_meta_flux_time_n6} the instantaneous flux as a function of time $t$ (right axis) in addition to the flow (accumulated flux at time $t$, left axis) are plotted. The time instances corresponding to Figs.~\ref{fig:n_6_wk_1_particle_0p03088}-\ref{fig:n_6_wk_1_particle_0p03888} are marked in Fig.~\ref{fig:case3_meta_flux_time_n6}.  The motion of the fluid particles near the third cilium under the influence of the velocity field caused by the ciliary motion is also shown.
It can be observed from Fig.~\ref{fig:case3_meta_flux_time_n6} that one beat cycle consists of six sub-beats, which correspond to  the traveling of the magnetic couple from one cilium to the next.  The traveling of the metachronal wave to the right can, for instance, be seen by looking at the cilia which exhibit the recovery stroke (i.e.~cilium 1 in Fig.~\ref{fig:n_6_wk_1_particle_0p03088}, cilium 2 in Fig.~\ref{fig:n_6_wk_1_particle_0p032}, etc).  The negative flow created by the cilia during their recovery stroke is overcome by the flow due to the effective stroke of the rest of the cilia; this leads to a  vortex formation near the cilia exhibiting their recovery stroke. As a result, the negative flow is completely obstructed for most of the time during the recovery stroke.
It can be observed from Fig.~\ref{fig:case3_meta_flux_time_n6} that no flux (right axis) is transported in the negative direction, and that the flow (left axis) continuously increases during each sub-beat. 
Moreover, the increase in the flow during each sub-beat is similar (see Fig.~\ref{fig:case3_meta_flux_time_n6}). Thus, the total flow per beat cycle (left axis of Fig.~\ref{fig:case3_meta_flux_time_n6})  is the sum of the flows generated during each sub-beat (i.e.~flow per beat = 6$\times$flow generated during one sub-beat). Therefore, it is sufficient to analyse the fluid flow during  one sub-beat.

\begin{figure}
\begin{minipage}{0.6\textwidth }
  \centering
       \subfigure[\ $t=t_\text{beat}/6$]{\includegraphics[scale=1.]{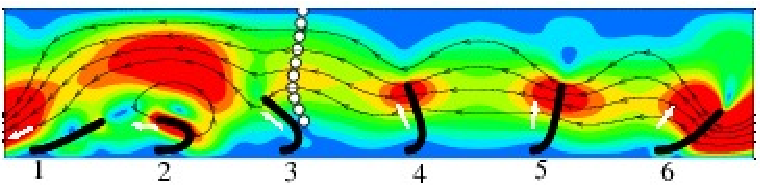}\label{fig:n_6_wk_1_1}}\\
    \subfigure[\ $t=0.25 t_\text{beat}$]{\includegraphics[scale=1.]{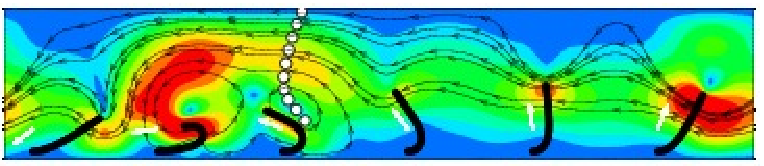}\label{fig:n_6_wk_1_2}}\\
   \subfigure[\ $t=0.316 t_\text{beat}$]{\includegraphics[scale=1.]{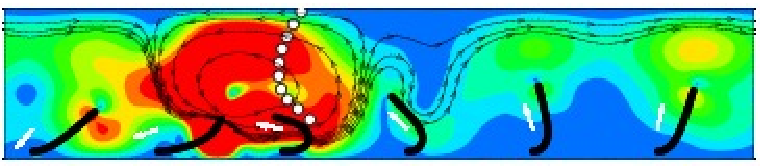}\label{fig:n_6_wk_1_3}}\\
\subfigure[\ $t=2t_\text{beat}/6$]{\includegraphics[scale=1.0]{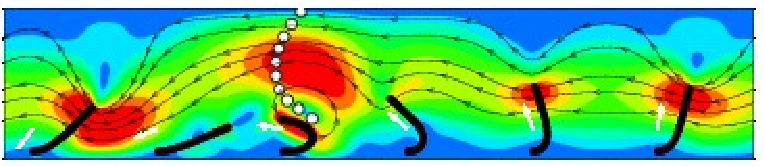}\label{fig:n_6_wk_1_4}}\\
\end{minipage}
\begin{minipage}{0.1\textwidth}
  \subfigure[]{\includegraphics{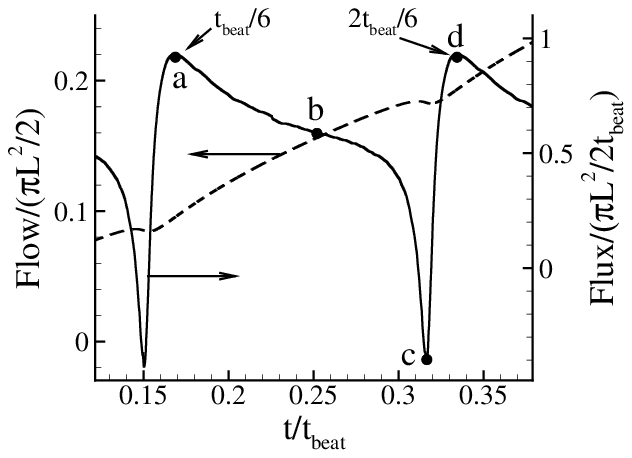}\label{fig:case3_meta_flux_time_n6_zoomed_in}}
\includegraphics{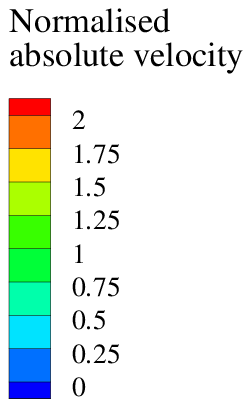}
\end{minipage}
    \caption{(Color online) (a)-(d) Snapshots for the out-of-phase motion of cilia between time instances of Figs.~\ref{fig:n_6_wk_1_particle_0p032} and \ref{fig:n_6_wk_1_particle_0p0335} for $\Delta\phi=\pi/6$ ($n=12$) with the wave moving to the  right (antiplectic metachrony) for $a/L=1.67$.   The contours represent the absolute velocity normalised with $L/t_\text{beat}$.  The direction of the velocity is represented by streamlines. The white circles represent fluid particles. The applied magnetic field at each cilium is represented by the white arrows. (e) Instantaneous  flux  (right axis) and flow (left axis) as a function of time with the instances (a)-(d) duly marked.  }
     \label{fig:case3_meta_particle_tracking_short}
\end{figure}

In the following, we analyse the fluid motion and the resulting flow during the second sub-beat. The velocity profiles at different instants  of this sub-beat are shown in Figs.~\ref{fig:n_6_wk_1_1}-\ref{fig:n_6_wk_1_4}. The corresponding flow and the flux generated are shown in Fig.~\ref{fig:case3_meta_flux_time_n6_zoomed_in}.
At $t_\text{beat}/6$, the third cilium starts its recovery stroke and the particles near the top boundary are driven by the positive flow created by cilia 4, 5 and 6 (see Fig.~\ref{fig:n_6_wk_1_1}). At this instant, as only one cilium is exhibiting a recovery stroke, the flux created by the cilia is maximum (see instant `a' in Fig.~\ref{fig:case3_meta_flux_time_n6_zoomed_in}). In Fig.~\ref{fig:n_6_wk_1_2}, the third cilium also has begun its recovery stroke and now the negative flow caused by both the second and third cilia is opposed by the effective stroke of the other cilia. The high velocity of the second cilium during its recovery stroke decreases the flux caused by the other cilia (see instant `b' in Fig.~\ref{fig:case3_meta_flux_time_n6_zoomed_in}).  When the third cilium is half-way through its recovery stroke (see  Fig.~\ref{fig:n_6_wk_1_3}), the second cilium is about to finish its recovery, which generates a large velocity, due to the whip-like action \cite[]{khaderi}, to the right. Now, the position of the third cilium is such, that it opposes the negative flow caused by the second cilium. This leads to a strong vortex formation near the second and third cilia, with only a small flux in the direction of the recovery stroke (to the right). The small negative flux caused by the whip-like motion of the second cilium can be seen by the instant marked `c' in Fig.~\ref{fig:case3_meta_flux_time_n6_zoomed_in}, causing a momentary decrease in the flow. The vortex imparts a high velocity in the direction of the effective stroke to the particles away from the cilia. As the third cilium progresses further in its recovery stroke, the particles come under the influence of the flow due to the rest of cilia, which are now in different phases of their effective stroke (see Fig.~\ref{fig:n_6_wk_1_4}). Now, only the third cilium is in the recovery stroke; this again leads to a maximum value of the flux (similar to Fig.~\ref{fig:n_6_wk_1_1}). 
The key observation of Figs.~\ref{fig:case3_meta_particle_tracking} and \ref{fig:case3_meta_particle_tracking_short} is that the negative flow created during the recovery stroke of the cilia creates a local vortex due to the positive flow created by  other cilia. This shielding effect during the recovery stroke leads to a drastic increase in the net propulsion rate for  cilia beating out-of-phase, compared to synchronously beating cilia.

\begin{figure}
\centering
\subfigure[]{\includegraphics[width = 7 cm] {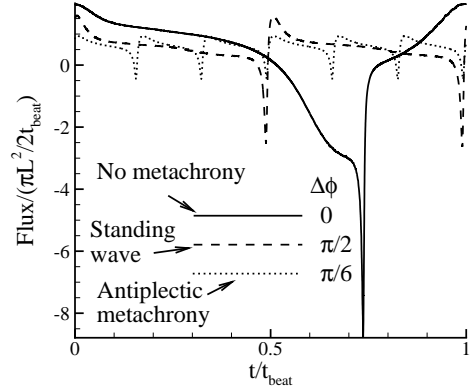}\label{fig:case3_meta_flux_with_time}}
\subfigure[]{\includegraphics[width = 7 cm] {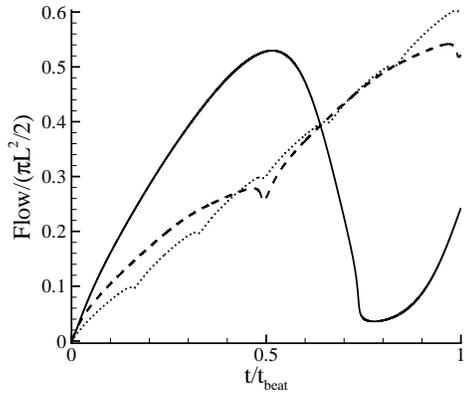}\label{fig:case3_meta_flow_with_time}}
    \caption{(a) Normalised fluid flux as a function of time for $a/L=1.67$ and different values of phase difference $\Delta\phi.$ (b) Normalised accumulated flow at any time $t$ during the beat cycle. }  
    
\end{figure}

Next, we analyse the instantaneous flux (Fig.~\ref{fig:case3_meta_flux_with_time}) and flow generated (Fig.~\ref{fig:case3_meta_flow_with_time}) as a function of time for different phase differences.
When the cilia  move synchronously ($\Delta\phi=0$), the flux  (see the solid line in Fig.~\ref{fig:case3_meta_flux_with_time}) is positive for approximately three-quarters of the time and strongly negative during the rest of the cycle. Consequently, the flow generated (see the solid line in Fig.~\ref{fig:case3_meta_flow_with_time}) increases during the effective stroke, but profoundly decreases when the recovery stroke takes place. This creates a large  fluctuation in the flow, with only a small net amount of fluid transported. Once the ciliary motion is metachronal, the  negative flux is very small compared to the positive flow (see the cases of a standing wave and antiplectic metachrony in Fig.~\ref{fig:case3_meta_flux_with_time}). This decreases the fluctuation in the flow generated, causing it to increase nearly monotonously during the beat cycle (see the dashed and dotted lines in Fig.~\ref{fig:case3_meta_flow_with_time}). We can clearly see that the flow at the end of the beat cycle ($t=t_\text{beat}$) for out-of-phase motion is significantly larger than the flow created by the synchronously beating cilia.   

\begin{figure}
	              \centering
      \subfigure[\ Area flow]{\includegraphics[width=7cm ]{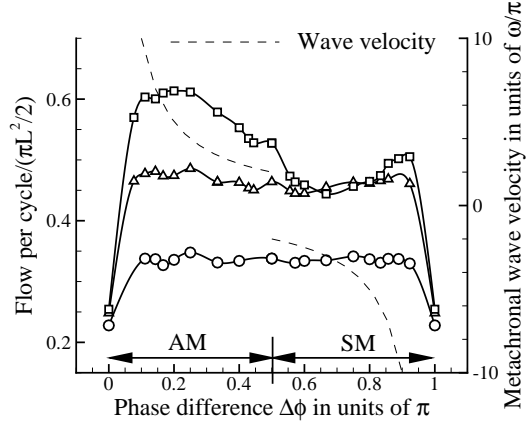}\label{fig:cilia_meta_results_flow}}
      \subfigure[\ Effectiveness]{\includegraphics[width=7cm ]{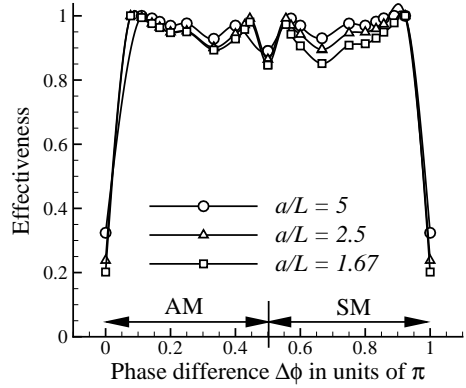}\label{fig:cilia_meta_results_effectiveness}}
\caption{Flow and effectiveness as a function of the phase difference $\Delta\phi$ for different inter-cilium spacings $a/L$. AM and SM refer to antiplectic metachrony (the wave direction is opposite to the direction of the effective stroke)  and symplectic metachrony (the wave direction and the effective stroke direction are the same), respectively. }
\label{fig:cilia_meta_results}
\end{figure}

\begin{figure}
	              \centering
      \subfigure[]{\includegraphics[width=7cm ]{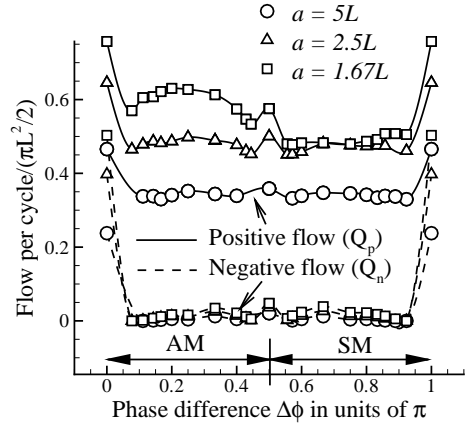}\label{fig:cilia_meta_results_flow_positive_negative}}
      \subfigure[]{\includegraphics[width=7cm ]{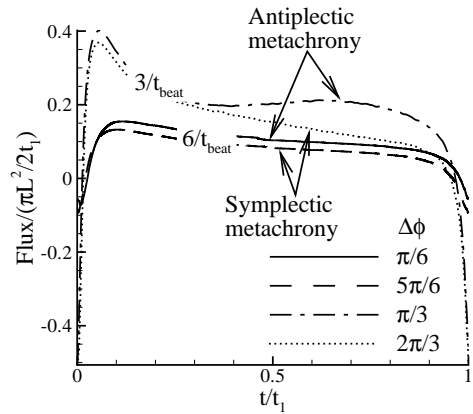}\label{fig:cilia_meta_flux_scaled_time}}
      \caption{ \subref{fig:cilia_meta_results_flow_positive_negative} Positive ($Q_p$) and  negative flow ($Q_n$) (see Fig.~\ref{fig:case3_meta_schematic}) created by the cilia corresponding to the results presented in Fig.~\ref{fig:cilia_meta_results}. \subref{fig:cilia_meta_flux_scaled_time} Flux vs time (scaled with the time $t_1$ taken by the magnetic couple to travel from one cilium to the next) for $a/L=1.67$ and different wave speeds.  }
\end{figure}

The fluid propelled and the corresponding effectiveness are plotted for different values of $\Delta\phi$ and $a/L$ in Fig.~\ref{fig:cilia_meta_results}. The metachronal wave velocity (Fig.~\ref{fig:case3_meta_apparent_wave_velocity}) is plotted as a function of $\Delta\phi$ and is shown using dashed lines in Fig.~\ref{fig:cilia_meta_results_flow}. As mentioned earlier, when the metachronal wave velocity is positive  an antiplectic metachrony (AM) results, and when the metachronal wave velocity is negative we get a symplectic metachrony (SM).
When all the cilia are moving synchronously ($\Delta\phi=0$ or $\pi$), the flow (normalised by $\pi L^2/2$)  will be approximately 0.22 for $a/L=5$. As the cilia density is increased by decreasing $a$ from $a/L=5$ to $a/L=1.67$, the viscous resistance per cilium decreases, which causes the normalised flow to increase to 0.25.  
When the cilia beat in-phase,  the effectiveness of fluid propulsion is very low, see Fig.~\ref{fig:cilia_meta_results_effectiveness}. 
The fluid propelled shows a substantial increase once the cilia start beating out-of-phase (Fig.~\ref{fig:cilia_meta_results_flow}).  
  When the cilia spacing is large ($a/L=5$ and $2.5$), the flow generated  remains approximately  constant  for all metachronal wave speeds. The increase in flow by decreasing the cilia spacing from $a/L=5$ to $a/L=2.5$ is much larger when the cilia beat out-of-phase compared to the increase when the cilia beat in-phase.
  However, when the cilia spacing is low ($a/L=1.67$),  we see a larger increase in the fluid flow when there is an antiplectic metachrony(AM)  compared to a symplectic metachrony (SM). 
  Also, the effectiveness sharply increases from around 0.3 (i.e., 30\% of the totally displaced fluid is converted into net flow) to 1 (fully unidirectional flow), see Fig.~\ref{fig:cilia_meta_results_effectiveness}.
 To analyse these trends a bit further, we plot the positive and negative flow ($Q_p$ and $Q_n$ in Fig.~\ref{fig:case3_meta_schematic}) created during a beat cycle for different phase differences in Fig.~\ref{fig:cilia_meta_results_flow_positive_negative}. It can be seen that the cilia do not create a negative flow when they beat out-of-phase for all cilia spacings, resulting in a unidirectional flow (effectiveness = 1).
 This reduction in negative flow is due to the shielding of flow during the recovery stroke caused by the effective flow of other cilia.
  It can also be noted that the positive flow is also reduced compared to in-phase beating, but the reduction is considerably less than the reduction in negative flow. Thus, the net flow increases as soon as the cilia start to beat out-of-phase (see Fig.~\ref{fig:cilia_meta_results_flow}). 
  It can be seen from Fig.~\ref{fig:cilia_meta_results_flow_positive_negative} that in the presence of metachronal waves when the cilia spacing is large ($a/L=5$), the fluid transported during the effective stroke remains nearly the same for all values of the wave velocities. 
For small cilia spacing ($a/L = 1.67$), however, the positive flow is maximal for antiplectic metachrony, which leads to a larger net flow for antiplectic metachrony compared to symplectic metachrony.

\begin{figure}
              \centering
    \subfigure[\ Antiplectic metachrony: wave travels to the right]{ \includegraphics[scale=1.2]{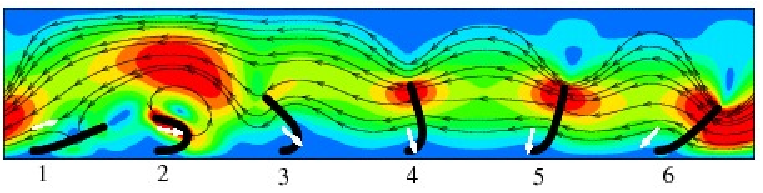}}
    \subfigure[\ Symplectic metachrony: wave travels to the left]{\includegraphics[scale=1.2]{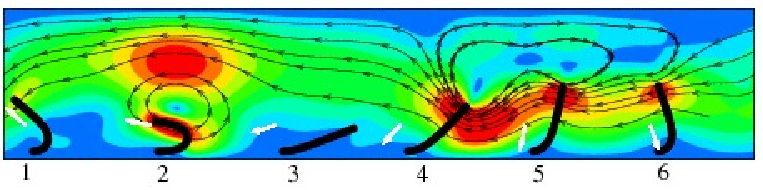}}

    \caption{(Colour online) Snapshots for antiplectic ($\Delta\phi=\pi/6$) and symplectic metachrony ($\Delta\phi=5\pi/6$)  for a wave speed of $6/t_\text{beat}$  cilia per second and cilia spacing $a/L=1.67$ at $t=0.1 t_1$ of Fig.~\ref{fig:cilia_meta_flux_scaled_time}.
 The contours represent the absolute velocity normalised with $L/t_\text{beat}$ (blue and red colours represent a normalised velocity of 0 and 2, respectively).  The direction of the velocity is represented by streamlines. The applied magnetic field is shown by the white arrows. }

\label{fig:case3_meta_am_sm_comparison_at_small_spacing}
\end{figure}

To understand the difference in  positive flow for opposite wave directions  for small inter-cilium spacing ($a/L=1.67$), we plot the flux as a function of time scaled with the time taken by the magnetic couple to travel from one cilium to the next $t_1$, for two different metachronal wave velocities ($3/t_\text{beat}$ and $6/t_\text{beat}$ cilia per second), see Fig.~\ref{fig:cilia_meta_flux_scaled_time}. The corresponding phase differences are also shown in the legend. It can be seen that the flux in the case of antiplectic metachrony is larger than the flux created by the symplectic metachrony for the same wave speed. 
This  difference in flux for opposite wave directions can be understood by analysing the velocity field corresponding to symplectic and antiplectic metachrony at time instances when the flux is maximum (see Fig.~\ref{fig:case3_meta_am_sm_comparison_at_small_spacing}). Figure \ref{fig:case3_meta_am_sm_comparison_at_small_spacing}(a) and \ref{fig:case3_meta_am_sm_comparison_at_small_spacing}(b) correspond to different phase differences ($\Delta\phi=\pi/6$  and $\Delta\phi=5\pi/6$, respectively) leading to a similar wave speed of $6/t_\text{beat}$  cilia per second (see also Fig.~\ref{fig:case3_meta_apparent_wave_velocity}).  The fifth cilium is in the peak of its effective stroke for both AM and SM. 
In the case of symplectic metachrony, the positive flow created by the fifth cilium is obstructed by the close proximity of the fourth cilium, which has just started its effective stroke.  As a result, we observe  the formation of a vortex. In the case of antiplectic metachrony, however, the position of the fourth cilium is such that the positive flow created by the fifth cilium is not obstructed. This leads to larger fluid flow in the positive direction, so that the net flow created by an antiplectic metachrony is larger than that created by its symplectic counterpart. 
 
Reports on metachrony and phase locking of beating cilia have appeared in the past \cite[]{ gauger_cilia, kim_netz, ShayGueron06101997, ShayGueron_energy}.
The main results are that  metachrony enhances flow compared to synchronously beating cilia \cite[]{kim_netz, gauger_cilia} and that antiplectic metachrony generates a higher flow rate than symplectic metachrony \cite[]{gauger_cilia}. 
\cite{kim_netz} analysed two cilia, which are driven by internal motors and are moving out-of-phase due to the hydrodynamic interaction. They have shown that the fluid propulsion increases, once the cilia start to beat with a phase difference, which is in agreement with our results.
Our results also agree with \cite{gauger_cilia}, where it is shown that  the fluid flow is larger in the case of  antiplectic metachrony than symplectic metachrony  when the cilia are close together. However, our results differ from \cite{gauger_cilia} in the sense that  we always see an enhancement in flow in the presence of  metachrony (compared to cilia beating in-phase)  irrespective of the direction and magnitude of the metachronal wave velocity. This is most likely due to the fact that the asymmetry in ciliary motion in our case is much higher. 
\cite{ShayGueron06101997} and \cite{ShayGueron_energy} have proposed that the evolution of the out-of-phase motion of cilia in Paramecia is due to hydrodynamic interactions between adjacent cilia leading to antiplectic metachrony. It is interesting to observe that the interplay between the internally-driven actuation and hydrodynamic interaction in nature results in antiplectic metachrony. Our results, and those of others \cite[]{gauger_cilia}, show that indeed antiplectic metachrony leads to larger flow than symplectic metachrony for small cilia spacings as typically seen in nature.

\section{Conclusions}\label{sec:conclusion_out-of-phase}

Using a numerical model we have studied the flow created by a two-dimensional array of plate-like artificial cilia as a function of the phase lag and spacing between neighbouring cilia. The flow per cycle and the effectiveness (which is a measure of the unidirectionality of flow) are considerably enhanced when the cilia start beating out-of-phase, as compared to synchronously beating cilia.  While the amount of flow enhancement depends on the inter-cilia spacing,  the effectiveness  is not significantly influenced. Metachrony is observed to completely knock-down the negative flow to zero due to the vortex formation caused by the shielding of the recovery stroke. Interestingly, we find that the enhancement is achieved even for small phase differences. The direction of travel of the metachronal wave is important only for small cilia spacing. In that case, the flow is larger for antiplectic metachrony compared to symplectic metachrony, which is related to the obstruction of the positive flow for symplectic metachrony. It is therefore beneficial if the magnetic actuation of the artificial cilia is designed such that it results in an antiplectic metachrony. Our results suggest that an antiplectic metachrony is adopted by the cilia on paramecia and in the respiratory system to maximize the fluid propelled. However, ciliary systems (such as on Opalina) that exhibit symplectic metachrony are also present in nature. It will be of interest to investigate what property is optimised by symplectic metachrony in these systems.

\section*{Acknowledgements}

This work is a part of the $\text{6}^\text{th}$ Framework European project 'Artic', under contract STRP 033274. We would also like to acknowledge fruitful discussions with Michiel Baltussen and Patrick Anderson.
\appendix
\section{Metachronal wave velocity}\label{sec:apparent_velocity}
The metachronal wave velocity is obtained by dividing the distance between two cilia with the time it takes for the magnetic couple to travel from a cilium to its neighbor. If the neighbor is to the right, then the wave  travels to the right, and when the neighbor is to the left, the wave travels to the left. The magnetic couple $N_{i}$ at any cilium $i$ is proportional to  $\sin\left( 2\omega t -    2 \phi_i \right) $, and travels with a phase velocity of $\omega/\Delta\phi$ (in number of cilia per second) to the right.

\begin{figure}
  \centering
\includegraphics[scale=1]{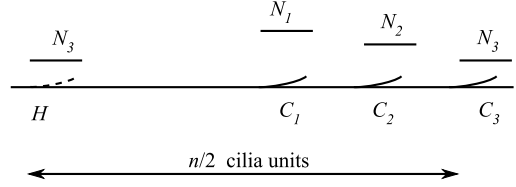}
\caption{Schematic diagram used to calculate the metachronal wave velocity.}
\label{fig:velocity_apparent_schematic_meta}
\end{figure}

In the schematic of Fig.~\ref{fig:velocity_apparent_schematic_meta}, three cilia $C_1$, $C_2$ and $C_3$ are depicted.  At any given instance of time, let the magnitude of the magnetic couple at $C_1$, $C_2$ and $C_3$ be  $N_1$, $N_2$ and $N_3$, respectively. The magnitude of the magnetic couple at the `periodic' cilium $H$, which is separated from $C_3$ by $n/2$ units, is also $N_3$.  The metachronal wave is said to have traveled to the right when the magnetic field at $C_2$ is $N_1$ after a time interval. Now, the distance traveled by the magnetic couple is 1 cilia spacing, and the time taken to travel this distance is $1/(\omega/\Delta\phi)$.  Therefore, the velocity of the magnetic couple  is $\omega/\Delta\phi$, in cilia units per second. The metachronal wave is said to have traveled to the left when the magnetic field at $C_2$ is equal to $N_3$ after an interval of time. As the applied magnetic couple travels to the right, this situation is possible when the magnetic couple at the periodic cilium $H$ travels to the cilium $C_2$. 
The time needed for the magnetic couple to travel from $H$ to $C_2$ is equal to $(n/2-1)/(\omega/\Delta\phi)$. However, the apparent distance travelled is one cilium spacing to the left (i.e.~from $C_3$ to $C_2$), so that the wave velocity is now $\omega/(\pi-\Delta\phi)$. The (apparent) metachronal wave velocity is now determined by the maximum of the two competing wave velocities: $\omega/\Delta\phi$ to the right and $\omega/(\pi-\Delta\phi)$ to the left.  As a  result,   the metachronal wave  velocity is equal to $\omega/\Delta\phi$ (i.e.~to the right) when $\omega/\Delta\phi>\omega/(\pi-\Delta\phi)$ (i.e.~$0<\Delta\phi<\pi/2$), and it is equal to  $-\omega/(\pi-\Delta\phi)$ (i.e.~to the left) when $\omega/\Delta\phi<\omega/(\pi-\Delta\phi)$ (i.e.~$\pi/2<\Delta\phi<\pi$), see Fig.~\ref{fig:case3_meta_apparent_wave_velocity}.

\section{Validation of the fluid-structure interaction model}\label{sec:benchmark}
\begin{figure}\centering
   \includegraphics[width=12 cm]{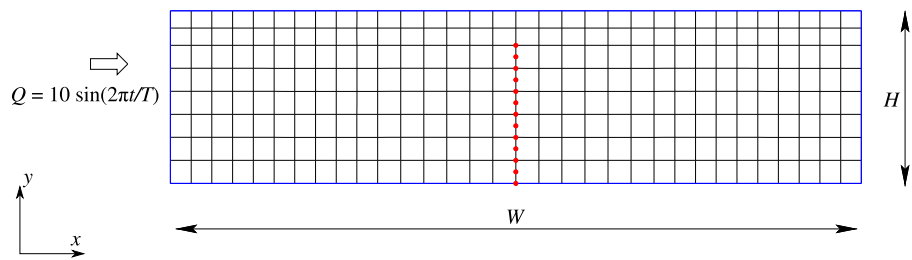}
   \caption{Coarsest mesh used for benchmarking.}\label{fig:baaijens_mesh}
\end{figure}

\begin{figure}\centering
   \subfigure[]{\includegraphics[width=52mm]{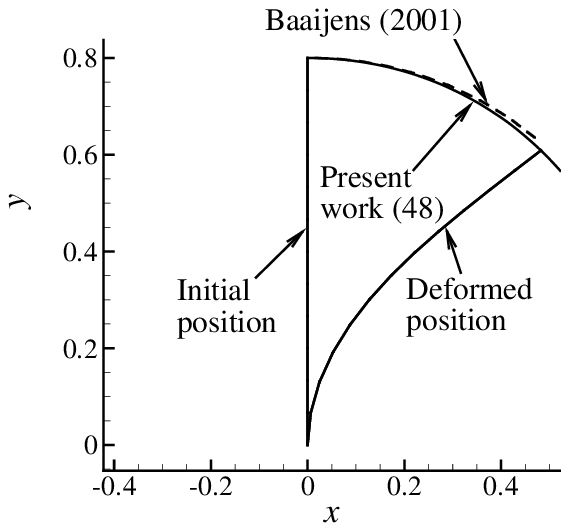}\label{fig:baaijens_xy}}
   \subfigure[]{\includegraphics[width=65mm]{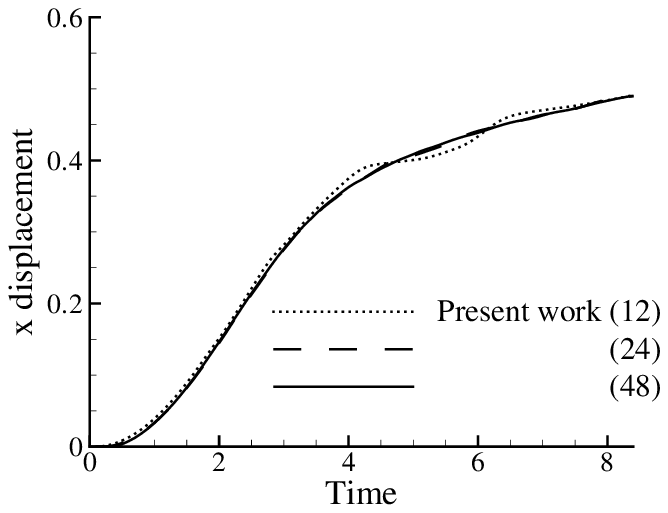}\label{fig:baaijens_x_implicit_time}}
   \caption{A cantilever subjected to a pulsating  flow: Comparison of solution obtained from the present work  with \cite{frank_baaijens}. \subref{fig:baaijens_xy} Comparison of the trajectory of the free end. The deformed and initial configurations  are also shown.  \subref{fig:baaijens_x_implicit_time} Comparison of the displacement of the free end as a function of time for various mesh refinements. The number in parenthesis  of the legend refer to the number of elements used to discretize the cantilever. }\label{fig:baaijens_comparison}
\end{figure}

\begin{figure}\centering
         \subfigure[]{\includegraphics[scale=2]{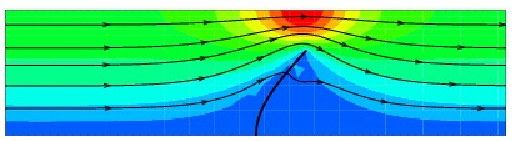}}
         \subfigure[]{\includegraphics[scale=2]{fig_13a.eps}}
         \subfigure[]{\includegraphics[scale=2]{fig_13a.eps}}
         \includegraphics[scale=1.2]{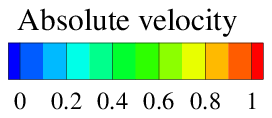}
	 \caption{Convergence of velocity field at a particular time instant with mesh refinement. The mesh used for Fig.~(a) is shown in Fig.~\ref{fig:baaijens_mesh}, where 12 beam elements are used. In Fig.~(b) and (c), 24 and 48 elements were used to discretize the cilia, while the fluid mesh was also refined proportionally. }
	 \label{fig:baaijens_vel_conv}
\end{figure}

To compare the performance of the present approach with a solution available in the literature we choose to study the deformation behavior of a cantilever beam under an imposed pulsating flow. This problem has been numerically solved by \cite{frank_baaijens} using the fictitious domain method in which the solid was discretized using continuum finite elements. The width $W$ is four times the height $H$ of the fluid domain. $H$ is taken to be unity. The length of the cilium is $0.8H$. The thickness of the cilium is $0.0212H$. The elastic modulus of the cilium and viscosity of the fluid were specified in dimensionless units to be $E=10^7$ and $\mu= 10$, respectively. The mesh used for the computation is shown in Fig.~\ref{fig:baaijens_mesh}. The dots represent the nodes of the Euler-Bernoulli beam element. The boundary conditions are as follows: the left and right boundaries are periodic. A pulsating flow of magnitude  $10\sin (2\pi t/T)$ is prescribed on the  left boundary, where $T$ is the time period which is taken to be sufficiently large to avoid inertia effects in the cilium. The bottom boundary is a no slip boundary. On the top boundary, the normal flow is constrained. The solution from our formulation is plotted along with the solution from  \cite{frank_baaijens} in Fig.~\ref{fig:baaijens_comparison}(a) in terms of the displacement of the free end of the cantilever.  It can been seen that the two solutions are in good agreement. In Fig.~\ref{fig:baaijens_comparison}(b), we plot the $x$ displacement of the free end of the beam as a function of time for different discretizations of the cilium (using 12, 24 and 48 beam elements). When the cilium mesh is refined, the fluid mesh is also refined proportionally, see also appendix C.   It can be seen that the displacements nicely converge as the mesh is refined. The convergence of the velocity field is also shown in Fig.~\ref{fig:baaijens_vel_conv}.

\section{Convergence of the numerical model}\label{sec:convergence}
\begin{figure}\centering
  \includegraphics[width=5 cm]{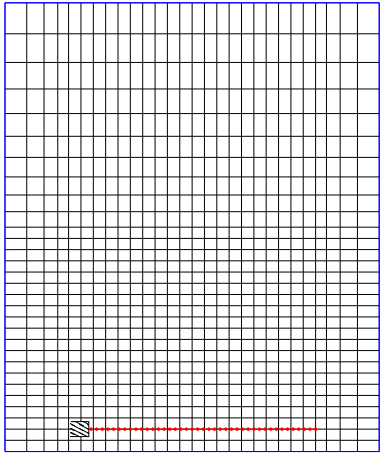}
  \caption{Discretization used for cilium and fluid. The cilium is discretized into 40 elements and the fluid domain of size $a \times H$ is divided into 28 $\times$ 30 elements. }
  \label{fig:conv_mesh}
\end{figure}

In this section, we report on the spatial and temporal convergence of the numerical method used in this paper. We use the case of synchronously beating cilia  ($\Delta\phi=0$) with an inter-cilia spacing of $a=1.67L$, for which the unit-cell consists of one cilium. As the deformed shape of the cilium is an outcome of the model, we compare the position of the free end for different temporal discretizations.  The mesh used to discretize the cilium  and the fluid domain is shown in Fig.~\ref{fig:conv_mesh} for the case when the cilium is divided into 40 cilia elements and the fluid is divided into 28 $\times$ 30 elements.

\begin{figure}\centering
  \subfigure[]{\includegraphics[width=6.57 cm]{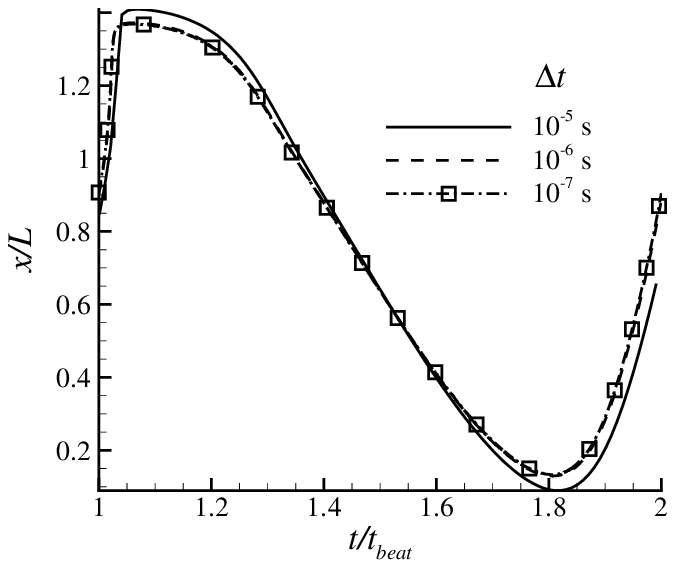}}
  \subfigure[]{\includegraphics[width=6.57 cm]{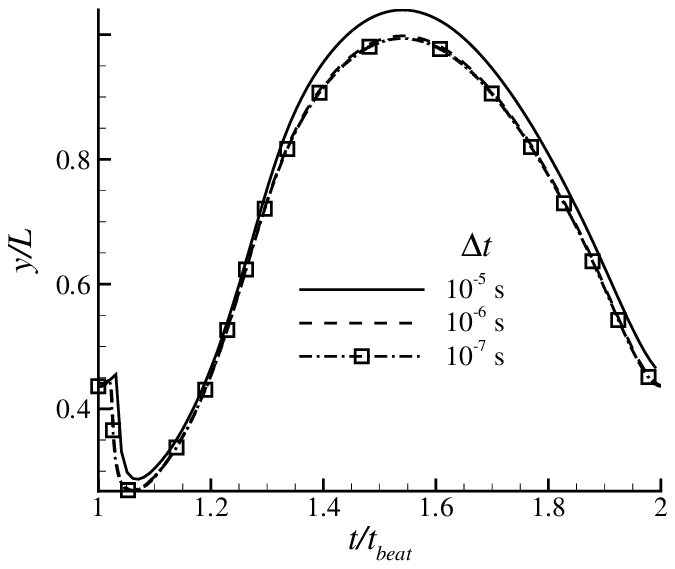}}
  \subfigure[]{\includegraphics[width=6.57 cm]{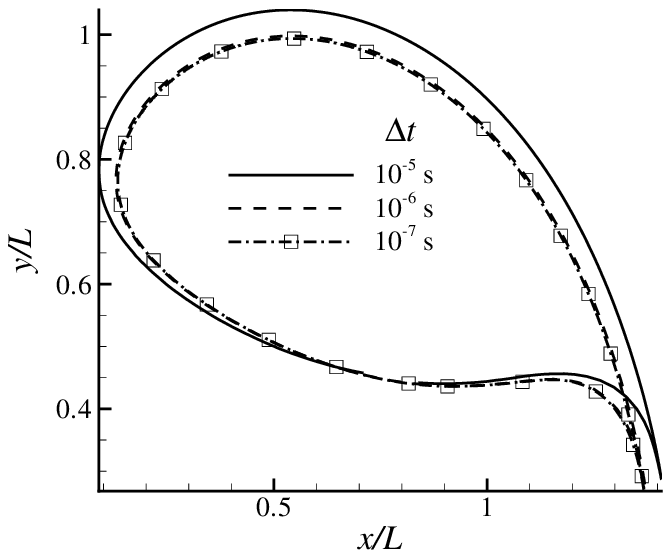}}
  \caption {(a)-(b) Temporal convergence: Position of the tip of the cilium as a function of time for different time increments $\Delta t$. (c) The trajectory of the free end of the cilium for different time increments. The cilium is divided into 40 elements.  }
  \label{fig:temp_conv}
\end{figure}

\begin{figure}\centering
  \subfigure[]{\includegraphics[width=6.57 cm]{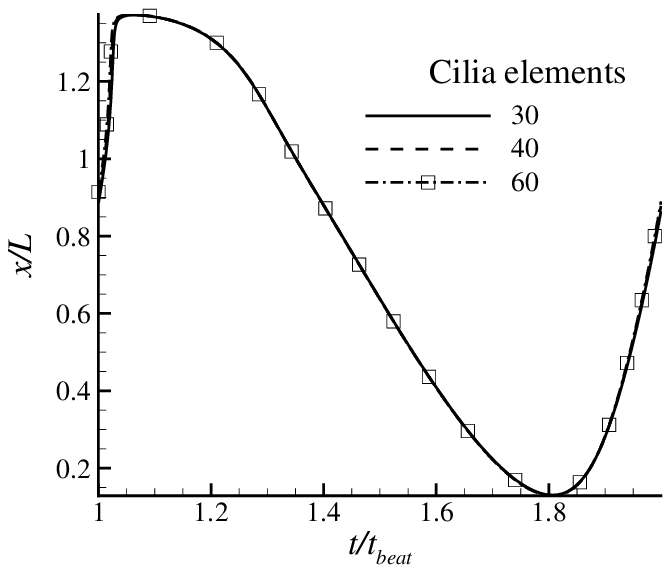}}
  \subfigure[]{\includegraphics[width=6.57 cm]{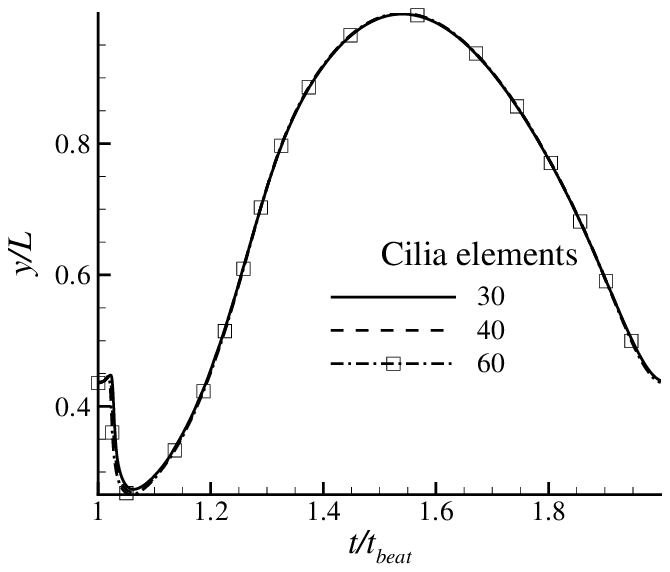}}
  \subfigure[]{\includegraphics[width=6.57 cm]{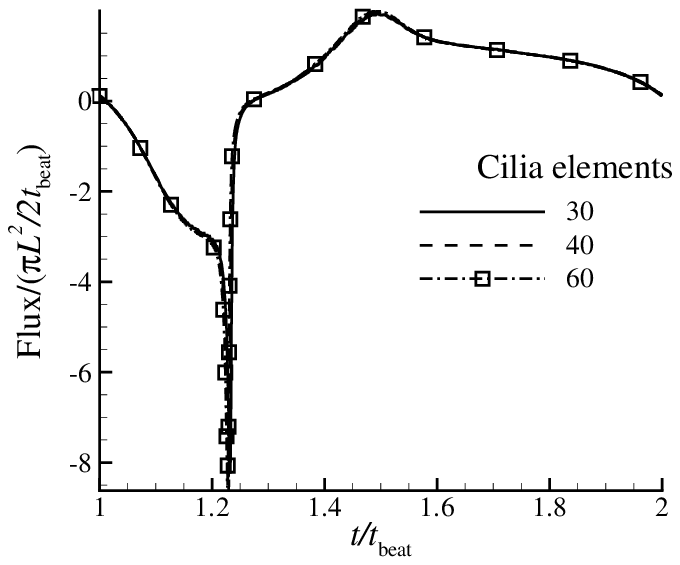}}
  \caption {(a)-(b) Position of the tip of the cilium as a function of the time for different spatial discretizations.  (c) Flux as a function of time for different spatial discretizations.   }
  \label{fig:spatial_conv}
\end{figure}

The position of the tip of the cilium as a function of time and its trajectory for different time increments is shown in Fig.~\ref{fig:temp_conv} (a)-(c). The time increment has to be small enough to capture the fast whip-like recovery stroke. It can be seen that a time increment of 1 $\mu$s is sufficient for temporal convergence. This time step of 1 $\mu$s is used to study the spatial convergence and the results are shown in Fig.~\ref{fig:spatial_conv}. The number of elements on the cilium as well as the fluid are changed proportionally when the mesh is changed. In the following the spatial discretization is defined in terms of the number of elements used to discretise the cilium; i.e., 30 cilia elements correspond to a fluid mesh of 21 $\times$ 23 and 60 cilia elements correspond to a fluid mesh of 42 $\times$ 45.  It can be seen that the results for these discretizations have fully converged as shown for the position of the free end of the cilium and the flux as a function of time. 


\begin{thebibliography}{44}
\expandafter\ifx\csname natexlab\endcsname\relax\def\natexlab#1{#1}\fi

\bibitem[Annabattula {\em et~al.\/}(2010)Annabattula, Huck \& Onck]{ratna_jmps}
{\sc Annabattula, R.~K., Huck, W. T.~S. \& Onck, P.R.} 2010 Micron-scale
  channel formation by the release and bond-back of pre-stressed thin films: A
  finite element analysis. {\em Journal of the Mechanics and Physics of
  Solids\/} {\bf 58}, 447--465.

\bibitem[Baaijens(2001)]{frank_baaijens}
{\sc Baaijens, Frank P.~T.} 2001 A fictitious domain/mortar element method for
  fluid-structure interaction. {\em International Journal for Numerical Methods
  in Fluids\/} {\bf 35}~(7), 743--761.

\bibitem[Belardi {\em et~al.\/}(2010)Belardi, Schorr, Prucker, Wells, Patel \&
  Ruhe]{belardi_tolouse}
{\sc Belardi, J., Schorr, N., Prucker, O., Wells, S., Patel, V. \& Ruhe, J.}
  2010 Fabrication of articial rubber cilia by photolithography. In {\em Second
  European Conference on Microfluidics - paper no. 112\/}.

\bibitem[Blake(1971{\natexlab{{\em a\/}}})]{blake_envelope_1971}
{\sc Blake, J.~R.} 1971{\natexlab{{\em a\/}}} Infinite models for ciliary
  propulsion. {\em Journal of Fluid Mechanics\/} {\bf 49}~(02), 209--222.

\bibitem[Blake(1971{\natexlab{{\em b\/}}})]{blake_spherical_envelope}
{\sc Blake, J.~R.} 1971{\natexlab{{\em b\/}}} A spherical envelope approach to
  ciliary propulsion. {\em Journal of Fluid Mechanics\/} {\bf 46}~(01),
  199--208.

\bibitem[Blake(1972)]{blake_sublayer_1972}
{\sc Blake, J.~R.} 1972 A model for the micro-structure in ciliated organisms.
  {\em Journal of Fluid Mechanics\/} {\bf 55}~(01), 1--23.

\bibitem[Blake \& Sleigh(1974)]{blake_sleigh_review}
{\sc Blake, J.~R. \& Sleigh, M.~A.} 1974 Mechanics of ciliary locomotion. {\em
  Biological Reviews\/} {\bf 49}, 85--125.

\bibitem[Brennen \& Winet(1977)]{brennen_cilia_flagella}
{\sc Brennen, Christopher \& Winet, Howard} 1977 Fluid mechanics of propulsion
  by cilia and flagella. {\em Annual Review of Fluid Mechanics\/} {\bf 9},
  339--398.

\bibitem[Chen {\em et~al.\/}(2003)Chen, Ma, Tan \& Guan]{lingxin}
{\sc Chen, Lingxin, Ma, Jiping, Tan, Feng \& Guan, Yafeng} 2003 Generating
  high-pressure sub-microliter flow rate in packed microchannel by
  electroosmotic force: potential application in microfluidic systems. {\em
  Sensors and Actuators B: Chemical\/} {\bf 88}, 260--265.

\bibitem[Cook {\em et~al.\/}(2001)Cook, Malkus, Plesha, Malkus \&
  Plesha]{cook_fem}
{\sc Cook, Robert~D., Malkus, D.S., Plesha, M.E., Malkus, David~S. \& Plesha,
  Michael~E.} 2001 {\em Concepts and Applications of Finite Element
  Analysis\/}. John Wiley and Sons.

\bibitem[Dauptain {\em et~al.\/}(2008)Dauptain, Favier \& Bottaro]{Dauptain}
{\sc Dauptain, A., Favier, J. \& Bottaro, A.} 2008 Hydrodynamics of ciliary
  propulsion. {\em Journal of Fluids and Structures\/} {\bf 24}~(8), 1156 --
  1165, unsteady Separated Flows and their Control.

\bibitem[Evans {\em et~al.\/}(2007)Evans, Shields, Carroll, Washburn, Falvo \&
  Superfine]{evans_cilia}
{\sc Evans, B.~A., Shields, A.~R., Carroll, R.~Lloyd, Washburn, S., Falvo,
  M.~R. \& Superfine, R.} 2007 Magnetically actuated nanorod arrays as
  biomimetic cilia. {\em Nano Letters\/} {\bf 7}~(5), 1428--1434.

\bibitem[Fahrni {\em et~al.\/}(2009)Fahrni, Prins \& van
  IJzendoorn]{3dexpt_cilia}
{\sc Fahrni, Francis, Prins, Menno W.~J. \& van IJzendoorn, Leo~J.} 2009
  Micro-fluidic actuation using magnetic artificial cilia. {\em Lab on a
  Chip\/} {\bf 9}, 3413 -- 3421.

\bibitem[Gauger {\em et~al.\/}(2009)Gauger, Downton \& Stark]{gauger_cilia}
{\sc Gauger, E.~M., Downton, M.~T. \& Stark, H.} 2009 Fluid transport at low
  {R}eynolds number with magnetically actuated artificial cilia. {\em The
  European Physical Journal E\/} {\bf 28}, 231--242.

\bibitem[Gueron \& Levit-Gurevich(1999)]{ShayGueron_energy}
{\sc Gueron, S. \& Levit-Gurevich, K.} 1999 {Energetic considerations of
  ciliary beating and the advantage of metachronal coordination}. {\em
  Proceedings of the National Academy of Sciences of the United States of
  America\/} {\bf 96}~(22), 12240--12245.

\bibitem[Gueron {\em et~al.\/}(1997)Gueron, Levit-Gurevich, Liron \&
  Blum]{ShayGueron06101997}
{\sc Gueron, S., Levit-Gurevich, K., Liron, N. \& Blum, J.~J.} 1997 {Cilia
  internal mechanism and metachronal coordination as the result
  of hydrodynamical coupling}. {\em Proceedings of the National Academy of
  Sciences of the United States of America\/} {\bf 94}~(12), 6001--6006.

\bibitem[Jeon {\em et~al.\/}(2000)Jeon, Dertinger, Chiu, Choi, Stroock \&
  Whitesides]{JeonN.L._la000600b}
{\sc Jeon, N.L., Dertinger, S.K.W., Chiu, D.T., Choi, I.S., Stroock, A.D. \&
  Whitesides, G.M.} 2000 Generation of solution and surface gradients using
  microfluidic systems. {\em Langmuir\/} {\bf 16}~(22), 8311--8316.

\bibitem[Khaderi {\em et~al.\/}(2009)Khaderi, Baltussen, Anderson, Ioan, den
  Toonder \& Onck]{khaderi}
{\sc Khaderi, S.~N., Baltussen, M. G. H.~M., Anderson, P.~D., Ioan, D., den
  Toonder, J. M.~J. \& Onck, P.~R.} 2009 Nature-inspired microfluidic
  propulsion using magnetic actuation. {\em Physical Review E\/} {\bf 79}~(4),
  046304.

\bibitem[Khaderi {\em et~al.\/}(2010)Khaderi, Baltussen, Anderson, den Toonder
  \& Onck]{khaderi_inertia}
{\sc Khaderi, S.~N., Baltussen, M. G. H.~M., Anderson, P.~D., den Toonder, J.
  M.~J. \& Onck, P.~R.} 2010 {T}he breaking of symmetry in microfluidic
  propulsion driven by artificial cilia. {\em Physical Review E\/} {\bf 82},
  027302.

\bibitem[Kim \& Netz(2006)]{kim_netz}
{\sc Kim, Y.~W. \& Netz, R.~R.} 2006 Pumping fluids with periodically beating
  grafted elastic filaments. {\em Physical Review Letters\/} {\bf 96}~(15),
  158101.

\bibitem[Kinosita \& Murakami(1967)]{kinosita_review}
{\sc Kinosita, H. \& Murakami, A.} 1967 Control of ciliary motion. {\em
  Physiological Reviews\/} {\bf 47}, 53--82.

\bibitem[Laser \& Santiago(2004)]{0960-1317-14-6-R01}
{\sc Laser, D~J \& Santiago, J~G} 2004 A review of micropumps. {\em Journal of
  Micromechanics and Microengineering\/} {\bf 14}~(6), R35--R64.

\bibitem[Liron(1978)]{liron_cilia_between_plates}
{\sc Liron, N.} 1978 Fluid transport by cilia between parallel plates. {\em
  Journal of Fluid Mechanics\/} {\bf 86}~(04), 705--726.

\bibitem[{v}an Loon {\em et~al.\/}(2006){v}an Loon, Anderson \& van~de
  Vosse]{vanloon}
{\sc {v}an Loon, R., Anderson, P.~D. \& van~de Vosse, F.~N.} 2006 A
  fluid-structure interaction method with solid-rigid contact for heart valve
  dynamics. {\em Journal of Computational Physics\/} {\bf 217}, 806--823.

\bibitem[Malvern(1977)]{malvern}
{\sc Malvern, Lawrence~E.} 1977 {\em Introduction to the Mechanics of a
  Continuous Medium\/}. Prentice-Hall.

\bibitem[Mitran(2007)]{mitran}
{\sc Mitran, S.~M.} 2007 Metachronal wave formation in a model of pulmonary
  cilia. {\em Computers and Structures\/} {\bf 85}, 763–774.

\bibitem[Niedermayer {\em et~al.\/}(2008)Niedermayer, Eckhardt \&
  Lenz]{niedermayer}
{\sc Niedermayer, Thomas, Eckhardt, Bruno \& Lenz, Peter} 2008 Synchronization,
  phase locking, and metachronal wave formation in ciliary chains. {\em Chaos:
  An Interdisciplinary Journal of Nonlinear Science\/} {\bf 18}~(3), 037128.

\bibitem[van Oosten {\em et~al.\/}(2009)van Oosten, Bastiaansen \&
  Broer]{printed_cilia}
{\sc van Oosten, Casper~L., Bastiaansen, Cees W.~M. \& Broer, Dirk~J.} 2009
  Printed artificial cilia from liquid-crystal network actuators modularly
  driven by light. {\em Nature Materials\/} {\bf 8}, 677 -- 682.

\bibitem[Qian {\em et~al.\/}(2009)Qian, Jiang, Gagnon, Breuer \&
  Powers]{qian_syn}
{\sc Qian, Bian, Jiang, Hongyuan, Gagnon, David~A., Breuer, Kenneth~S. \&
  Powers, Thomas~R.} 2009 Minimal model for synchronization induced by
  hydrodynamic interactions. {\em Phys. Rev. E\/} {\bf 80}~(6), 061919.

\bibitem[van Rijsewijk(2006)]{Rijsewijk}
{\sc van Rijsewijk, L.} 2006 Electrostatic and magnetic microactuation of
  polymer structures for fluid transport. Master's thesis, Eindhoven University
  of Technology.

\bibitem[Roper {\em et~al.\/}(2006)Roper, Dreyfus, Baudry, Fermigier, Bibette
  \& Stone]{CambridgeJournals:431278}
{\sc Roper, Marcus, Dreyfus, R\'{e}mi, Baudry, Jean, Fermigier, M., Bibette, J.
  \& Stone, H.~A.} 2006 On the dynamics of magnetically driven elastic
  filaments. {\em Journal of Fluid Mechanics\/} {\bf 554}~(-1), 167--190.

\bibitem[Satir \& Sleigh(1990)]{satir_sale_annual_review}
{\sc Satir, P \& Sleigh, M~A} 1990 The physiology of cilia and mucociliary
  interactions. {\em Annual Review of Physiology\/} {\bf 52}~(1), 137--155,
  pMID: 2184754.

\bibitem[Schilling {\em et~al.\/}(2002)Schilling, Kamholz \&
  Yager]{SchillingE.A._ac015640e}
{\sc Schilling, E.A., Kamholz, A.E. \& Yager, P.} 2002 Cell lysis and protein
  extraction in a microfluidic device with detection by a fluorogenic enzyme
  assay. {\em Analytical Chemistry\/} {\bf 74}~(8), 1798--1804.

\bibitem[Schorr {\em et~al.\/}(2010)Schorr, Belardi, Prucker, Wells, Patel \&
  Ruhe]{schorr_tolouse}
{\sc Schorr, N., Belardi, J., Prucker, O., Wells, S., Patel, V. \& Ruhe, J.}
  2010 Magnetically actuated polymer flap arrays mimicking artificial cilia. In
  {\em Second European Conference on Microfluidics - paper no. 105\/}.

\bibitem[Shields {\em et~al.\/}(2010)Shields, Fiser, Evans, Falvo, Washburn \&
  Superfine]{Shields}
{\sc Shields, A.~R., Fiser, B.~L., Evans, B.~A., Falvo, M.~R., Washburn, S. \&
  Superfine, R.} 2010 {Biomimetic cilia arrays generate simultaneous pumping
  and mixing regimes}. {\em Proceedings of the National Academy of Sciences\/}
  .

\bibitem[Sing {\em et~al.\/}(2010)Sing, Schmid, Schneider, Franke \&
  Alexander-Katz]{katz_walker}
{\sc Sing, Charles~E., Schmid, Lothar, Schneider, Matthias~F., Franke, Thomas
  \& Alexander-Katz, Alfredo} 2010 {Controlled surface-induced flows from the
  motion of self-assembled colloidal walkers}. {\em Proceedings of the National
  Academy of Sciences\/} {\bf 107}~(2), 535--540.

\bibitem[Smith {\em et~al.\/}(2008)Smith, Gaffney \&
  Blake]{smith_modelling_review}
{\sc Smith, D.J., Gaffney, E.A. \& Blake, J.R.} 2008 Modelling mucociliary
  clearance. {\em Respiratory Physiology and Neurobiology\/} {\bf 163},
  178–188.

\bibitem[Smith {\em et~al.\/}(2007)Smith, Gaffney \&
  Blake]{smith_discrete_cilia}
{\sc Smith, D.~J., Gaffney, E.~A. \& Blake, J.~R.} 2007 Discrete cilia
  modelling with singularity distributions: Application to the embryonic node
  and the airway surface liquid. {\em Bulletin of Mathematical Biology\/} {\bf
  69}, 1477--1510.

\bibitem[Tecplot(2008)]{tecplot}
{\sc Tecplot} 2008 {T}ec360 user manual.

\bibitem[den Toonder {\em et~al.\/}(2008)den Toonder, Bos, Broer, Filippini,
  Gillies, de~Goede, Mol, Reijme, Talen, Wilderbeek, Khatavkar \&
  Anderson]{den_toonder}
{\sc den Toonder, Jaap, Bos, Femke, Broer, Dick, Filippini, Laura, Gillies,
  Murray, de~Goede, Judith, Mol, Titie, Reijme, Mireille, Talen, Wim,
  Wilderbeek, Hans, Khatavkar, Vinayak \& Anderson, Patrick} 2008 Artificial
  cilia for active micro-fluidic mixing. {\em Lab on a Chip\/} {\bf 8}~(4),
  533--541.

\bibitem[Vilfan \& J\"ulicher(2006)]{vilfan_synchronization}
{\sc Vilfan, Andrej \& J\"ulicher, Frank} 2006 Hydrodynamic flow patterns and
  synchronization of beating cilia. {\em Phys. Rev. Lett.\/} {\bf 96}~(5),
  058102.

\bibitem[Vilfan {\em et~al.\/}(2010)Vilfan, Potocnik, Kavcic, Osterman,
  Poberaj, Vilfan \& Babic]{self_assembles_cilia}
{\sc Vilfan, Mojca, Potocnik, Anton, Kavcic, Blaz, Osterman, Natan, Poberaj,
  Igor, Vilfan, Andrej \& Babic, Dusan} 2010 Self-assembled artificial cilia.
  {\em Proceedings of the National Academy of Sciences\/} {\bf 107},
  1844--1847.

\bibitem[West {\em et~al.\/}(2002)West, Karamata, Lillis, Gleeson, Alderman,
  Collins, Lane, Mathewson \& Berney]{jonathan_west}
{\sc West, Jonathan, Karamata, Boris, Lillis, Brian, Gleeson, James~P.,
  Alderman, John, Collins, John~K., Lane, William, Mathewson, Alan \& Berney,
  Helen} 2002 Application of magnetohydrodynamic actuation to continuous flow
  chemistry. {\em Lab Chip\/} {\bf 2}, 224--230.

\bibitem[Zeng {\em et~al.\/}(2002)Zeng, Chen, Santiago, Chen, Zare, Tripp, S.
  \& Frechet]{shulin}
{\sc Zeng, S., Chen, C., Santiago, J.~G., Chen, J., Zare, R.~N., Tripp, J.~A.,
  S., F. \& Frechet, J. M.~J.} 2002 Electroosmotic flow pumps with polymer
  frits. {\em Sensors and Actuators B: Chemical\/} {\bf 82}~(2-3), 209 -- 212.

\end{thebibliography}
\end{document}